\documentclass[twocolumn,showpacs,preprintnumbers,amsmath,amssymb]{revtex4}
\usepackage{graphicx}
\usepackage{bm}% bold math
\begin{document}

%\preprint{APS/123-QED}

\title{Scaling properties of pyrex and silicon surfaces blasted with sharp particles}% Force line breaks with \\

\author{Z. Moktadir$^*$}
 \affiliation{School of Electronics and Computer Science,
Southampton University, Southampton, SO17 1BJ, United Kingdom}%Lines break
 \email{zm@ecs.soton.ac.uk}
 \author{H. Wensink}
 \affiliation{Transducers Science and Technology MESA+ Research Institute University of Twente, The Netherlands}
 \author{M. Kraft}
\affiliation{School of Electronics and Computer Science,
Southampton University, Southampton, SO17 1BJ, United Kingdom}

\begin{abstract}
The blasting  of brittle materials with sharp particles is an important fabrication technology in many industrial processes.
In particular, for micro-systems, it allows the production of devices with feature sizes down to few tens of microns.
An important parameter of this process is the surface roughness
of post-blasted surfaces. In this work the scaling properties of Pyrex glass and silicon surfaces after bombardment
with alumina particles is investigated. The targets were bombarded at normal incidence using alumina particles
with two different average sizes, $29\mu m$ and $9\mu m$, respectively. This investigation indicates that
the resulting surfaces have multifractal properties. Applying multifractal detrended fluctuation analysis (MFDFA)
allowed us to determine the singularity spectrum of the surfaces. This spectrum did not depend on the target material
 or on the size of the particles. Several parameters quantifying relevant quantities were determined.
We argue that for scales below $5 \mu m$, fracture processes are dominant while at large scales
long range correlations are responsible for the multifractal behaviour.

\end{abstract}

\maketitle
\section{Introduction}
Powder blasting technology is among several techniques used in the
micro-machining~\cite{Slikkerveer2000,Belloy99,Veenstra2001} of
devices on silicon and other materials. With such abrasive
techniques, one can achieve high erosion rates; higher than those
that can be obtained with conventional dry, or wet etching
processes; such as plasma etching or chemical etching. In the
field of micro-electromechanical-systems (MEMS), powder blasting
has already been used for the fabrication of inertial
sensors\cite{Belloy99}, peristaltic micro-pumps\cite{Veenstra2001}
and  miniaturized capillary electrophoresis
chips\cite{Schlautmann2001}.  Erosion with sharp particles is also
a widely used technique in aerospace and automotive industries.
Because of its involvement in many applications, it is important
to investigate the surface morphology of blasted surfaces resulting
from erosion since the performance of many devices will depend on the surface roughness. Such investigation  will also give an insight into the physical
mechanisms at work during  the bombardment of materials with particles. The mechanisms involved in the erosion of
brittle materials with sharp particles
 have been the subject of several studies. Several
 models have been developed \cite{Slikkerveer98, Chen2005} that are based on simple static indentation theory. These
 models empirically relate the erosion rate to the material's properties such as the
fracture toughness, the hardness and the Young modulus of the material.  These models simply state that an indentation force
is generated by the impact of bombarding particles which, in turn,
results in the formation of crack patterns. Some cracks penetrate
the material radially away from the surface into the bulk
material; others will nucleate and form a lateral ring parallel to
the surface\cite{Lawn75, Marshall82, Cook90}. The radial cracking
process was extensively studied \cite{Evans76_a, Lawn98} and was
used as a method for materials toughness
measurement\cite{Evans76_b}. Lateral cracks are responsible for the removal of material
 in abrasive and wear experiments on  brittle
materials\cite{Hagan78,Finnie95,Lawn93}.\\
 In separate studies, substantial work was dedicated to
the understanding of fracture surfaces~\cite{Bouchaud97} in
brittle materials. These studies focus on the scaling properties of fracture surfaces
resulting from an applied load on the
material. These surfaces are found to be
self-affine, i.e. the
root-mean-square surface fluctuations, averaged over a distance $L$
follows the scaling relation\cite{Family}:
\begin{equation}\label{scalingAnsatz}
    r \sim L^\alpha
\end{equation}
where the scaling exponent $\alpha$ is often called  the roughness
or the Hurst exponent. In a variety of materials, the
value of $\alpha$ was found to be approximately 0.8 over two or three
decades of scaling range. For this reason, it was conjectured
to be universal i.e. independent of the material (ductile or
brittle), the fracture mode  and the fracture
toughness~\cite{Bouchaud97,Maly92, Schmittbuhl94}. However, this
universality was questioned since the discovery of a second
exponent at the nanometer scale\cite{Milman93, Milman94,McAnulty92}. The value of this
second exponent is significantly smaller than 0.8 and close to
0.5. To explain this fact, it was proposed that the fracture
front could be imagined as a line moving through a random medium. Thus, the evolution of the
crack front can be described by a local nonlinear Langevin
equation \cite{Bouchaud93, Ertas92, Ertas93} which
predicts a crossover between two regimes corresponding to
$\alpha=0.5$ at small scales and to $\alpha=0.75$ at large scale.
In the framework of this model, the crossover length decreases
rapidly with the crack speed which was also predicted by numerical
simulations\cite{Nakano95}.\\ The present work is dedicated to the
study of the scaling properties of surfaces resulting from a
 bombardment by sharp particles. Pyrex glass (borosilicate glass) and silicon were the materials used in this investigation.
We will investigate the effect of the material and the size of
 the bombarding particles on those properties.
\section{Experiments} 
\begin{figure}[]
  \includegraphics[width=10 cm]{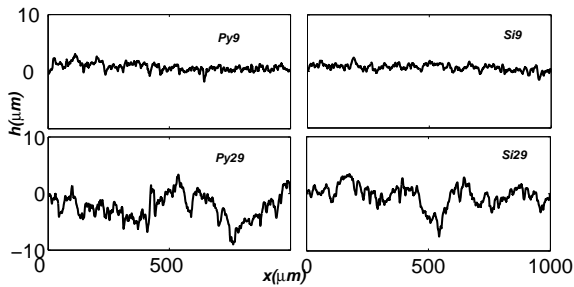}
  \caption{Surface profiles after blasting of Pyrex and silicon with particles of size $\delta= 9 \mu m$ (Py9 and Si9),
   and with particles of size $\delta= 29 \mu m$ (Py29 and Si29).}\label{profiles}
\end{figure}

In the experiments carried out, surfaces are exposed to a
directed particle jet, which results in mechanical material
removal. The particles are accelerated towards the target with a
high-pressure air flow through a circular nozzle (with a diameter
of 1.5 $\mu m$). The particles hit the target under normal
incidence, with an average speed of 290 m/s, in a ventilated box. A lateral movement of the target ensures an
evenly etched surface. The average diameter of the bombarding alumina
particles was $9 \mu m$ and $29 \mu m$, respectively. We performed measurements on
one-dimensional cuts of blasted surface using a mechanical surface profiler(Sloan
Dektak II), over a length of 1 mm. Each scanned profile is made of 8000 data points. Typical profiles obtained after
blasting Pyrex and silicon surfaces by alumina particles are shown
in figure \ref{profiles}.
Note the difference in the roughness amplitude in the two cases
 corresponding to different particle sizes. Large amplitudes are
 obtained when the targets are blasted by large particles. The
 typical rms surface roughness is $\sigma= 0.7 \mu m$ and $\sigma= 0.65 \mu m$ for Pyrex and silicon respectively, when $9 \mu m$ particles
 are used.  In table \ref{table1}, measured values of the surface rms
 roughness are shown, as a function of the bombarding particle
 sizes and the target material. Samples
bombarded with particles having a diameter $\delta$ are
denoted $X\delta$, where X denotes the material's symbol. For
example $Si29$ means silicon bombarded with particles having a
diameter of $29 \mu m $.
 \begin{table}

 \begin{ruledtabular}
 \begin{tabular}{lcr}
    Particles size & $9 \mu m$ & $29 \mu m$ \\ \hline
   Pyrex & $0.7\pm 0.02$ $\mu m$ &  $2.2 \pm 0.02$ $\mu m$\\
  Silicon & $0.65\pm 0.01$ $\mu m$ &  $2.3 \pm 0.01$ $\mu m$\\
 \end{tabular}
 \end{ruledtabular}\caption{Values of the rms roughness as a function of the particles size and the target
   material.}\label{table1}
 \end{table}\\
\section{Scaling properties of blasted surfaces}
\subsection{Global scaling exponents}
\begin{figure}[!h]
  \includegraphics[width=9 cm]{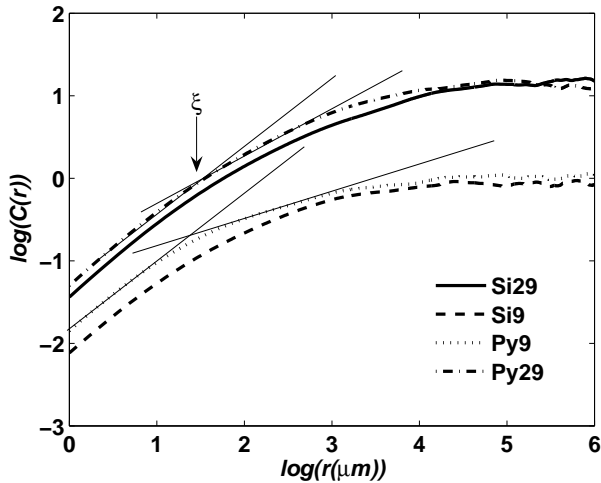}
  \caption{Plot of the logarithm of the height-height correlation function versus the logarithm of the distance, for Pyrex glass and silicon
  blasted with alumina particles with sizes of 29 and 9 microns. This plot reveals the existence of a cross-over length scale $\xi$ separating two
  scaling regions having two different global roughness exponents.}\label{hhcorr}
\end{figure}

We first investigate the global scaling behaviour of the blasted
surfaces by calculating the height-height correlation function
defined as :
\begin{equation}\label{rmseq}
    C(r)=\left<\left(h(x+r)-h(x)\right)^2\right>^{1/2}
\end{equation}
where the outer brackets mean the average over all positions $x$ and
$h(x)$ defines the surface profile. This function measures how
correlated  two points are on the profiles at a distance $r$ of
each other. If the surface is self-affine then $C(r) \propto
r^\alpha$, over the scaling range. Here, $\alpha$ is the roughness
exponent to be determined.  We computed $C(r)$ for one dimensional cuts and averaged the result
over a total of ten scanned profiles. In figure \ref{hhcorr}
we plot $C(r)$ for $2 \mu m < r < 60 \mu m$. We notice the
existence of a cross-over, occurring at a characteristic length
scale $\xi$, separating small and large scales with two different
roughness exponents, i.e. for $r<\xi$, $C(r) \propto r^{\alpha_1}$
and for $r>\xi$, $C(r) \propto r^{\alpha_2}$. The linear fit of
$log(C)$ {\it vs} $log(r)$ in the scaling region, determines the
value of the roughness exponent $\alpha_1$ for small scales and
$\alpha_2$ at large scale. These values are summarized in table
\ref{table2}, along with the  characteristic length scale $\xi$.

\begin{table}
\begin{tabular}{|c|c|c|c|c|}
  \hline
  Sample & $\alpha_1$ & $\alpha_2$ & $\xi(\mu m)$ \\
  \hline
  Si9 & 0.77 $\pm$ 0.02 & 0.4 $\pm$ 0.01 & 4.8  \\
  \hline
  Si29 & 0.81 $\pm$ 0.01 & 0.55 $\pm$ 0.002 & 4.7 \\
  \hline
  Py9 & 0.74 $\pm$ 0.02 & 0.32 $\pm$ 0.003 & 4.7 \\
  \hline
  Py29 & 0.8 $\pm$ 0.02  &   0.53 $\pm$  0.04 &5.2   \\
  \hline

\end{tabular}\caption{Values of the roughness exponents and the crossover length $\xi$ for the four samples.
}\label{table2}
\end{table}

\subsection{Multifractal  properties and singularity spectrum}
\begin{figure}
  \includegraphics[width=9 cm]{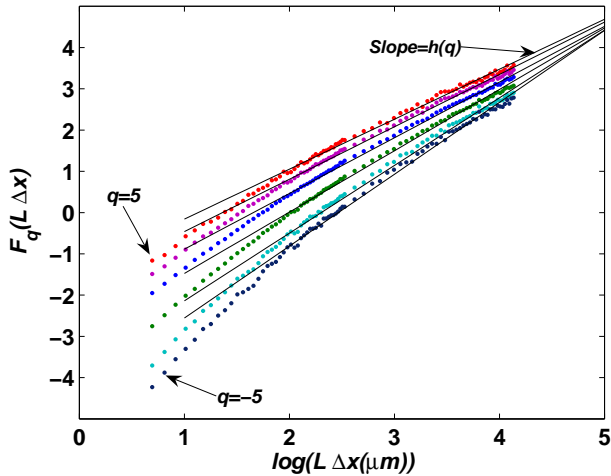}
  \caption{The logarithm of the fluctuation function $F_q(L \Delta x)$ as a function of the logarithm of the distance $d=L\Delta x$, for Py29 sample, for q=-5, -3, -1,
   1, 3 and 5. The straight lines
  are the regression fits to the data at large scale. The slopes of the lines determine $h(q)$.}\label{Fq_vs_q}
\end{figure}

\begin{figure}
  \includegraphics[width=9 cm]{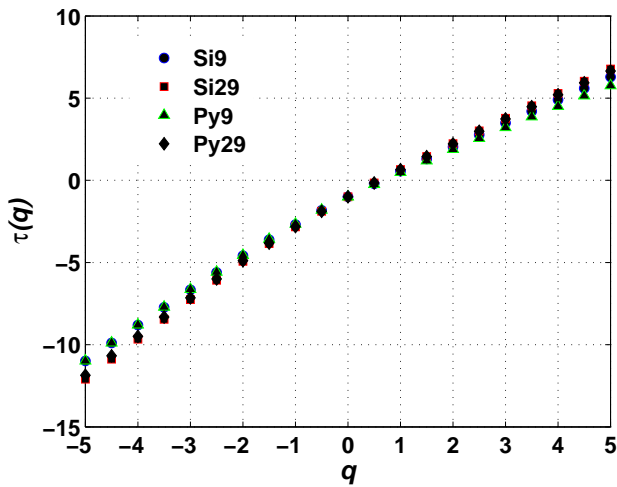}
  \caption{The q dependence of $\tau(q)=qh(q)-1$ for the four samples.}\label{tau_vs_q}
\end{figure}

\begin{figure}
  \includegraphics[width=10 cm]{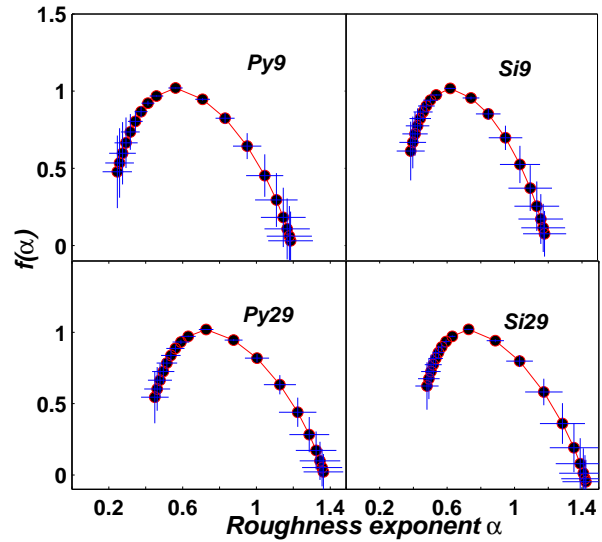}
  \caption{The singularity spectrum for the four samples. In each case, the spectrum was averaged
  over 12 profiles containing 8000 points.}\label{mfspectrum}
\end{figure}

The global analysis performed above can only reveal the existence of two roughness exponents. In many situations however, rough profiles exhibit a range
of roughness exponents. Such profiles are called multifractal while
profiles exhibiting a single roughness exponent are called monofractal\cite{Mandelbrot2}. Our aim is to determine the full range of local
roughness exponents for the blasted surfaces using multifractal analysis. To do so, we
characterized the profiles by computing the so-called {\it
singularity spectrum}, which determines the
distribution of the whole range of local roughness exponents. We used the
multifractal detrended fluctuation analysis
(MFDFA) \cite{Kantelhardt02}. This method has become popular thanks
to its simplicity and its easy computer implementation. Other
methods exist such as the wavelet transform modulus maxima
method(WTMM) \cite{Muzy91} but the MFDFA has several
advantages \cite{Oswiecimka06}. The MFDFA is the extension of the
detrended fluctuation(DFA) \cite{Peng} method which was used to
compute the roughness exponent of monofractal signals and for the
identification of long range correlations in non-stationary time
series \cite{Kantelhardt02}. MFDFA is an efficient tool to eliminate
undesirable trends in fluctuations. This method applied to our experimental data can be summarized as
follow: Given a profile $h(i)$, $i=1,...N$, we compute the
integrated profile,
\begin{equation}\label{IntegratedProfile}
    H(i)=\sum_{i=1}^{N}\left(h(i)-<h>\right)
\end{equation}
where N is the number of data points and $<h>$ is the mean height of the profile. The the whole
profile is subdivided into $M_L=N/L$ non-overlapping segments of
length $L$ (here $L$ is the number of data points in each segment). Since $M_L$ is not always an integer, some data will
be ignored during this procedure. To take them into account, the
subdivision is performed from both ends of the profile, which
results into $2M_L$ segments. In each segment $n$ the polynomial trend $P_n$ is subtracted
from the data. This polynomial is determined by a least square fit to the data in each segment. Polynomials of degrees higher than 1 can be used,
 corresponding to MFDFA2, MFDFA3, etc. After detrending in each
segment, the variance of the result is calculated :
\begin{equation}\label{variance}
    F^2(n,L)=\frac{1}{L}\sum_{j=1}^{j=L}[H((n-1)L+j)-P_n(j)]^2
\end{equation}
 This expression is then averaged over all segments $n$ and
 the value of the q$^{th}$-order fluctuation function is calculated \cite{Kantelhardt02}:
\begin{equation}\label{partition}
            F_q(L)=\left(\frac{1}{2M_L}\sum_{n=1}^{2M_L}F^2(n,L)^{q/2}\right)^{1/q}
\end{equation}
Here $q$ is a real number. For a fractal profile $F_q(L)$ follows
a power law relation at large scales i.e.:
\begin{equation}\label{powerlaw}
    F_q(L) \sim L^{h(q)}
\end{equation}
The exponent $h(q)$ is called the "generalised Hurst exponent" \cite{Kantelhardt02}. For a monofractal profile, $h(q)=const$ while for a
multifractal profile $h$ is a function of $q$. For positive values
of $q$, $h(q)$ describes the scaling behaviour of the segments with
large fluctuations, while for negative values of $q$, $h(q)$
describes the scaling behaviour of the segments with small
fluctuations. The singularity spectrum $f(\alpha)$ is calculated by
performing the Legendre Transform\cite{Feder88} of $(q,\tau(q))$ with
$\tau(q)=qh(q)-1$, resulting in:
\begin{eqnarray}\label{Legendre}
  \alpha &=& \frac{d\tau(q)}{dq} \nonumber \\
  f(\alpha) &=& q\alpha-\tau(q)
\end{eqnarray}
The spectrum $f(\alpha)$ can be interpreted as the fractal
dimension of a subset of points in the profile characterized by
the singularity strength  $\alpha$(the local roughness exponent).
For a monofractal profile, $\alpha=\alpha_0$ and $f(\alpha)=1$,
where $\alpha_0$ is the roughness exponent of the profile.  The strength of the
multifractality of a profile can be characterized by the
difference between the maximum and minimum values of $\alpha$,
i.e. $\Delta\alpha=\alpha_{max}-\alpha_{min}$.\\ Figure
\ref{Fq_vs_q} shows the logarithm of the fluctuation function $(F_q(d))$ versus
the logarithm of the distance $d=L\Delta x$, where $\Delta x$ is the spatial increment of the profile,  for Py29 sample. Also shown is
the linear regression fit at large scale for each value of $q$.
We used a detrended polynomial of degree 1( MFDFA1), but the
result remains unchanged when polynomial of degree two and three
were used(MFDFA2,MFDFA3). We can see clearly that at large scale,
the fluctuation functions $F_q(L)$
 are straight lines in the double logarithmic plot, with different slopes indicating the presence of multifractality. In figure \ref{tau_vs_q} the plot
of $\tau(q)$ versus $q$ is shown for the four samples Py9, Py29,
Si9 and Si29. Note that for $|q|>5$, the function $\tau(q)$ coincides with
its asymptotic form which is a linear function of q \footnote{Note that the function $\tau(q)$ defined in the MFDFA formalism, is equivalent
to functions defined in standard multifractal formalism. For rigorous definitions and properties of $\tau(q)$ see for example \cite{cowles}. }.
 Hence, we choose $-5 \leq q \leq 5$. We notice clearly that $\tau(q)$ is a concave
function of function of $q$, typical of multifractal profiles. From
equations (\ref{Legendre}) we can estimate the singularity spectrum
for the four samples as shown in figure \ref{mfspectrum}.  This
spectrum is averaged over 12 profile containing 8000 points each.
In table \ref{table3} we summarize the values of the strength of
the multifractality $\Delta\alpha$ for the four samples.
\begin{table}
\begin{tabular}{|p{1.5cm}|p{1.5cm}|p{1.5cm}|p{1.5cm}|p{1.5cm}|}
  \hline
  Sample & Py9 & Si9 & Py29 & Si29 \\
  \hline
  $\Delta\alpha$ & 0.94 & 0.8 & 0.91 & 0.94 \\
  \hline
\end{tabular}\caption{Values of the multifractality strength $\Delta\alpha=\alpha_{min}-\alpha_{max}$ for different samples.
}\label{table3}
\end{table}
 \section{Discussion}
\begin{figure}
  \includegraphics[width=9 cm]{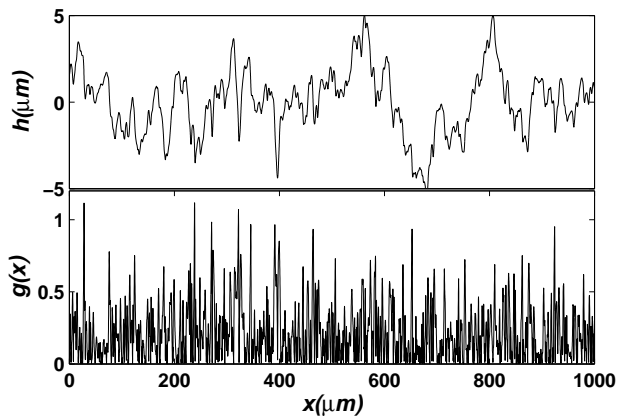}
  \caption{Top plot, the surface profile of Py29 sample with the corresponding local gradient (bottom plot) showing large fluctuations of the latter which constitutes
   the signature of intermittency. }\label{stepsize}
\end{figure}

\begin{figure}
  \includegraphics[width=9 cm]{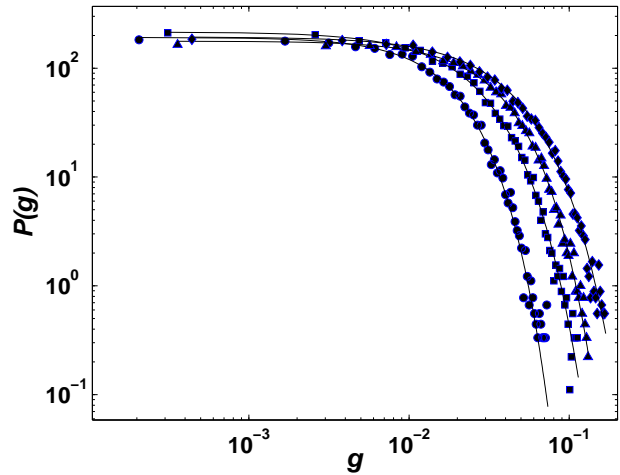}
  \caption{Local gradient distribution of the four samples Py9($\square$), Py29($\lozenge$), Si9($\bigcirc$) and Si29($\vartriangle$). Continuous lines are the fit to the stretched exponential
   distribution (\ref{strecheddistr}). }\label{strechedpdf}
\end{figure}

  The values of the two global roughness exponents were determined for
  all four samples Py9, Py29, Si9 and Si29 using the height-height
  correlation function. All samples show a crossover behavior
  corresponding to two values of the roughness exponent $\alpha_1$
  and $\alpha_2$ shown in table \ref{table2}. The value $\alpha_1$ is, within the error range, independent of the material or the size of impacting particles.
  For all analyzed samples, the small scale
  value is consistent with the approximate universal value of $0.8$ found for
  three dimensional
  fracture of brittle materials\cite{Bouchaud90,Maly92,Schmittbuhl95}. This provides evidence that at scales below the characteristic length $\xi \sim 5 \mu m$, the
  dominant mechanism in powder blasting is the fracture formation. This observation is in agreement with the static indentation
  theory\cite{Lawn75, Marshall82, Cook90} where the {\it dynamic load} of impacting particles creates a
  local load which increases the local stress resulting in the
  formation of lateral cracks, which are responsible for the
  material removal. In general, the universal value of the roughness exponent
  0.8 corresponding to three dimensional fracture surfaces is
  observed at large length scales (from 0.1 $\mu m$ to 1 $mm$, see reference \cite{Bouchaud97}). In contrast, the dynamic load of impacting particles induces a
  cross over to smaller values of the global roughness exponent, above 5 $\mu m$(
  see table \ref{table2}) which are 0.4, 0.32, 0.55 and 0.53 for
  Si9, Py9, Si29 and Py29, respectively.  The large scale values of
  the global roughness exponent obtained for Py9 and Si9 are smaller than
  those obtained for Si29 and Py29.  The effect of impacting particle's size
  is felt at large scales. Larger particles give larger roughness
  exponents, independent of the material being Silicon or Pyrex
  glass.\\
  We also showed that the surface generated by particles blasting
  share a common property of multifractality. The MFDFA method was
  used to uncover this property and to determine the singularity
  spectrum for each four samples. The strengths of multifractality represented by
  $\Delta\alpha$, as displayed in table \ref{table3}, are close to
  each other regardless of the size of the particles and the
  material. This multifractality  could be interpreted in terms of spatial
  intermittency. This concept was argued by Krug \cite{Krug94} to
  describe the scaling of surfaces generated by epitaxial growth models
  incorporating very limited atomic mobility leading to  a
  violent spatial intermittent effects  and multifractal  surfaces. This description was borrowed
  from fluid turbulence owing the similarity between Galilean invariance of turbulent fluids and translational invariance of interfaces\cite{Bohr92}.
  In our case, we highlight  large fluctuations by
  considering the step size  or the gradient
  at the position $x$,$g(x)=|\partial h/\partial x|$ for a profile $h(x)$. This quantity is the analog of energy dissipation $\varepsilon=(\partial v/\partial x)^2$,
  where $v$ is the local velocity of a turbulent fluid\cite{Frisch}. Figure \ref{stepsize} shows the local gradient $g(x)$ for Pyrex glass
  bombarded by $29 \mu m$-size particles, showing large gradient
  fluctuations which suggests spatial intermittency. In addition, we consider the distribution of the local gradient
  in analogy with local velocity gradient in fully
  developed turbulence, which is described by a stretched exponential
  distribution\cite{Bershadskii, Kailasnath92}. A very similar behaviour is characteristic of
  all four samples Py9, Py29, Si9 and Si29 as shown in figure
  \ref{strechedpdf},  where the local gradient distribution fits very
  well with the stretched exponential function:
  \begin{equation}\label{strecheddistr}
    P(g)=\frac{1}{\Omega}\exp(-ag^\gamma)
\end{equation}
The fit of the local gradient distribution to equation (\ref{strecheddistr}) gives the values of the stretching exponent $\gamma$, which are $1.41 \pm
0.05$, $1.23 \pm 0.03$, $1.37 \pm 0.05$ and $1.21 \pm 0.03$ for
Si9, Py9, Si29 and Py29 respectively.  The form of the local gradient distribution suggests that non-linearities must be present at large scale,
leading to  $h\rightarrow -h$ symmetry breaking. Indeed, the skewness $s$ of the profiles for all samples are non-zero and have
the values s $\simeq$ -0.4, -0.35, -0.28 and -0.4 for Si9, Si29, Py9 and Py29 respectively. We can now ask the question: What is  the origin of the
observed multifractality? It well known that multiplicative cascades \cite{Mandelbrot,Frisch} models generate processes known
to have intermittent behaviour and a multifractal character, mirroring the presence of intermittent fluctuations with long-range correlations.
 These long-range correlations are generated at  large scale by features that hierarchically cascade  their influence to smaller scales.
 To detect the presence of long range correlations in our surface profiles, we perform the following test\cite{Ivanov99}: we generated a
surrogate data set by shuffling the height data in each profile. The newly generated data set preserves the
distribution of the height but destroys the long range
correlations, which means that the surrogate profiles will exhibit a monofractal behavior, if the multifractality originates from the long-range correlations
 and not from the height distribution. We performed the MFDFA and found that the surrogate data set is
 monofractal with the roughness exponent $\alpha=0.5$ for all for samples.  Thus, the observed multifractal behaviour is a result of long range correlations since
 the shuffling procedure preserves the height distribution. The effect of long-range elastic interaction in developing long-range correlations was reported by some authors
  in the case of elastic chains driven in a quenched random pinning~\cite{Tanguy98}, or during crack propagation~\cite{Gao89}. Undoubtedly a detailed theoretical investigation
  is needed in order to determine the origin of the long-range correlations in sharp particle's bombardment of brittle materials.
\section{Conclusion}
In conclusion we conducted a detailed scaling analysis of  surfaces of two brittle materials, Silicon and Pyrex glass, after bombardment with alumina particles
of two different sizes $9 \mu m$ and $29 \mu m$. The bombardment results in  multifractal surfaces. This multifractality is common to all samples regardless of the
nature of the material or the size of the particles. We determined the corresponding singularity spectrum revealing a broad range of scaling exponents.
We argued that for scales below $5 \mu m$, fracture processes are dominant while at large scales
long range correlations are responsible for the observed multifractal behavior.
\begin{acknowledgments}
The authors would like to thank E. Bouchaud for the useful discussions.
\end{acknowledgments}
\bibliography{manuscript}

\begin{thebibliography}{44}
\expandafter\ifx\csname natexlab\endcsname\relax\def\natexlab#1{#1}\fi
\expandafter\ifx\csname bibnamefont\endcsname\relax
  \def\bibnamefont#1{#1}\fi
\expandafter\ifx\csname bibfnamefont\endcsname\relax
  \def\bibfnamefont#1{#1}\fi
\expandafter\ifx\csname citenamefont\endcsname\relax
  \def\citenamefont#1{#1}\fi
\expandafter\ifx\csname url\endcsname\relax
  \def\url#1{\texttt{#1}}\fi
\expandafter\ifx\csname urlprefix\endcsname\relax\def\urlprefix{URL }\fi
\providecommand{\bibinfo}[2]{#2}
\providecommand{\eprint}[2][]{\url{#2}}

\bibitem[{\citenamefont{Slikkerveer et~al.}(2000)\citenamefont{Slikkerveer,
  Bouten, and de~Haas}}]{Slikkerveer2000}
\bibinfo{author}{\bibfnamefont{P.}~\bibnamefont{Slikkerveer}},
  \bibinfo{author}{\bibfnamefont{P.}~\bibnamefont{Bouten}}, \bibnamefont{and}
  \bibinfo{author}{\bibfnamefont{F.}~\bibnamefont{de~Haas}},
  \bibinfo{journal}{Sensors. and Actuators} \textbf{\bibinfo{volume}{85}},
  \bibinfo{pages}{296} (\bibinfo{year}{2000}).

\bibitem[{\citenamefont{Belloy et~al.}(1999)\citenamefont{Belloy, Thurre,
  Walckiers, Sayah, and Gijs}}]{Belloy99}
\bibinfo{author}{\bibfnamefont{S.}~\bibnamefont{Belloy}},
  \bibinfo{author}{\bibfnamefont{E.}~\bibnamefont{Thurre}},
  \bibinfo{author}{\bibfnamefont{A.}~\bibnamefont{Walckiers}},
  \bibinfo{author}{\bibnamefont{Sayah}}, \bibnamefont{and}
  \bibinfo{author}{\bibfnamefont{M.}~\bibnamefont{Gijs}},
  \bibinfo{journal}{Eurosensors XIII proceeding, the Hague, Holland} p.
  \bibinfo{pages}{827} (\bibinfo{year}{1999}).

\bibitem[{\citenamefont{Veenstra et~al.}(2001)\citenamefont{Veenstra,
  Berenschot, Gardeniers, Sanders, Elwenspoek, and van~den
  Berg}}]{Veenstra2001}
\bibinfo{author}{\bibfnamefont{T.~T.} \bibnamefont{Veenstra}},
  \bibinfo{author}{\bibfnamefont{J.~W.} \bibnamefont{Berenschot}},
  \bibinfo{author}{\bibfnamefont{J.~G.~E.} \bibnamefont{Gardeniers}},
  \bibinfo{author}{\bibfnamefont{R.~G.~P.} \bibnamefont{Sanders}},
  \bibinfo{author}{\bibfnamefont{M.~C.} \bibnamefont{Elwenspoek}},
  \bibnamefont{and} \bibinfo{author}{\bibfnamefont{A.}~\bibnamefont{van~den
  Berg}}, \bibinfo{journal}{J. Electrochem. Soc.}
  \textbf{\bibinfo{volume}{148}}, \bibinfo{pages}{G68} (\bibinfo{year}{2001}).

\bibitem[{\citenamefont{Schlautmann et~al.}(2001)\citenamefont{Schlautmann,
  Wensink, Schasfoort, Elwenspoek, and Berg}}]{Schlautmann2001}
\bibinfo{author}{\bibfnamefont{S.}~\bibnamefont{Schlautmann}},
  \bibinfo{author}{\bibfnamefont{H.}~\bibnamefont{Wensink}},
  \bibinfo{author}{\bibfnamefont{R.~M.} \bibnamefont{Schasfoort}},
  \bibinfo{author}{\bibfnamefont{M.}~\bibnamefont{Elwenspoek}},
  \bibnamefont{and} \bibinfo{author}{\bibfnamefont{A.~V.~D.}
  \bibnamefont{Berg}}, \bibinfo{journal}{J. Micromech. Microeng.}
  \textbf{\bibinfo{volume}{11}}, \bibinfo{pages}{386} (\bibinfo{year}{2001}).

\bibitem[{\citenamefont{Slikkerveer et~al.}(1998)\citenamefont{Slikkerveer,
  Bouten, in't Veld, and Scholen}}]{Slikkerveer98}
\bibinfo{author}{\bibfnamefont{P.~J.} \bibnamefont{Slikkerveer}},
  \bibinfo{author}{\bibfnamefont{P.}~\bibnamefont{Bouten}},
  \bibinfo{author}{\bibfnamefont{F.}~\bibnamefont{in't Veld}},
  \bibnamefont{and} \bibinfo{author}{\bibfnamefont{H.}~\bibnamefont{Scholen}},
  \bibinfo{journal}{Wear} \textbf{\bibinfo{volume}{217}}, \bibinfo{pages}{237}
  (\bibinfo{year}{1998}).

\bibitem[{\citenamefont{Chen et~al.}(2005)\citenamefont{Chen, Hutchinson, and
  Evans}}]{Chen2005}
\bibinfo{author}{\bibfnamefont{X.}~\bibnamefont{Chen}},
  \bibinfo{author}{\bibfnamefont{J.~W.} \bibnamefont{Hutchinson}},
  \bibnamefont{and} \bibinfo{author}{\bibfnamefont{A.~G.} \bibnamefont{Evans}},
  \bibinfo{journal}{J. Am. Ceram. Soc.} \textbf{\bibinfo{volume}{88}},
  \bibinfo{pages}{1233} (\bibinfo{year}{2005}).

\bibitem[{\citenamefont{Lawn and Fuller}(1975)}]{Lawn75}
\bibinfo{author}{\bibfnamefont{B.~R.} \bibnamefont{Lawn}} \bibnamefont{and}
  \bibinfo{author}{\bibfnamefont{E.~R.} \bibnamefont{Fuller}},
  \bibinfo{journal}{J. Mater. Sci.} \textbf{\bibinfo{volume}{10}},
  \bibinfo{pages}{2016} (\bibinfo{year}{1975}).

\bibitem[{\citenamefont{Marshall et~al.}(1982)\citenamefont{Marshall, Lawn, and
  Evans}}]{Marshall82}
\bibinfo{author}{\bibfnamefont{D.~B.} \bibnamefont{Marshall}},
  \bibinfo{author}{\bibfnamefont{B.~R.} \bibnamefont{Lawn}}, \bibnamefont{and}
  \bibinfo{author}{\bibfnamefont{A.~G.} \bibnamefont{Evans}},
  \bibinfo{journal}{J. Am. Ceram. Soc.} \textbf{\bibinfo{volume}{65}},
  \bibinfo{pages}{561} (\bibinfo{year}{1982}).

\bibitem[{\citenamefont{Cook and Pharr}(1990)}]{Cook90}
\bibinfo{author}{\bibfnamefont{R.~F.} \bibnamefont{Cook}} \bibnamefont{and}
  \bibinfo{author}{\bibfnamefont{G.~M.} \bibnamefont{Pharr}},
  \bibinfo{journal}{J. Am. Ceram. Soc.} \textbf{\bibinfo{volume}{73}},
  \bibinfo{pages}{787} (\bibinfo{year}{1990}).

\bibitem[{\citenamefont{Evans and Wilshaw}(1976)}]{Evans76_a}
\bibinfo{author}{\bibfnamefont{A.~G.} \bibnamefont{Evans}} \bibnamefont{and}
  \bibinfo{author}{\bibfnamefont{T.~R.} \bibnamefont{Wilshaw}},
  \bibinfo{journal}{Acta Mater.} \textbf{\bibinfo{volume}{24}},
  \bibinfo{pages}{939} (\bibinfo{year}{1976}).

\bibitem[{\citenamefont{Lawn}(1998)}]{Lawn98}
\bibinfo{author}{\bibfnamefont{B.~R.} \bibnamefont{Lawn}}, \bibinfo{journal}{J.
  Am. Ceram. Soc.} \textbf{\bibinfo{volume}{81}}, \bibinfo{pages}{1977}
  (\bibinfo{year}{1998}).

\bibitem[{\citenamefont{Evans and Charles}(1976)}]{Evans76_b}
\bibinfo{author}{\bibfnamefont{A.~G.} \bibnamefont{Evans}} \bibnamefont{and}
  \bibinfo{author}{\bibfnamefont{E.~A.} \bibnamefont{Charles}},
  \bibinfo{journal}{J. Am. Ceram. Soc.} \textbf{\bibinfo{volume}{59}},
  \bibinfo{pages}{371} (\bibinfo{year}{1976}).

\bibitem[{\citenamefont{Hagan and Swain}(1978)}]{Hagan78}
\bibinfo{author}{\bibfnamefont{J.~T.} \bibnamefont{Hagan}} \bibnamefont{and}
  \bibinfo{author}{\bibfnamefont{M.~V.} \bibnamefont{Swain}},
  \bibinfo{journal}{J. Phys. D: Appl. Phys.} \textbf{\bibinfo{volume}{11}},
  \bibinfo{pages}{2091} (\bibinfo{year}{1978}).

\bibitem[{\citenamefont{Finnie}(1995)}]{Finnie95}
\bibinfo{author}{\bibfnamefont{I.}~\bibnamefont{Finnie}},
  \bibinfo{journal}{Wear} \textbf{\bibinfo{volume}{186}}, \bibinfo{pages}{1}
  (\bibinfo{year}{1995}).

\bibitem[{\citenamefont{Lawn}(1993)}]{Lawn93}
\bibinfo{author}{\bibfnamefont{B.}~\bibnamefont{Lawn}},
  \emph{\bibinfo{title}{Fracture of Brittle solids 2nd edn}}
  (\bibinfo{publisher}{Cambridge University Press}, \bibinfo{year}{1993}).

\bibitem[{\citenamefont{Bouchaud}(1997)}]{Bouchaud97}
\bibinfo{author}{\bibfnamefont{E.}~\bibnamefont{Bouchaud}},
  \bibinfo{journal}{J. Phys.: Condens. Matter} \textbf{\bibinfo{volume}{9}},
  \bibinfo{pages}{4319} (\bibinfo{year}{1997}).

\bibitem[{\citenamefont{Family and Viscek}(1991)}]{Family}
\bibinfo{author}{\bibfnamefont{F.}~\bibnamefont{Family}} \bibnamefont{and}
  \bibinfo{author}{\bibfnamefont{T.}~\bibnamefont{Viscek}},
  \emph{\bibinfo{title}{Dynamics of Fractal Surfaces}}
  (\bibinfo{publisher}{World Scientific, Singapore,}, \bibinfo{year}{1991}).

\bibitem[{\citenamefont{M{\aa}l{\o}y et~al.}(1992)\citenamefont{M{\aa}l{\o}y,
  Hansen, Hinrichsen, and Roux}}]{Maly92}
\bibinfo{author}{\bibfnamefont{K.~J.} \bibnamefont{M{\aa}l{\o}y}},
  \bibinfo{author}{\bibfnamefont{A.}~\bibnamefont{Hansen}},
  \bibinfo{author}{\bibfnamefont{E.~L.} \bibnamefont{Hinrichsen}},
  \bibnamefont{and} \bibinfo{author}{\bibfnamefont{S.}~\bibnamefont{Roux}},
  \bibinfo{journal}{Phys. Rev. Lett.} \textbf{\bibinfo{volume}{68}},
  \bibinfo{pages}{213} (\bibinfo{year}{1992}).

\bibitem[{\citenamefont{Schmittbuhl et~al.}(1994)\citenamefont{Schmittbuhl,
  Roux, and Berthaud}}]{Schmittbuhl94}
\bibinfo{author}{\bibfnamefont{J.}~\bibnamefont{Schmittbuhl}},
  \bibinfo{author}{\bibfnamefont{S.}~\bibnamefont{Roux}}, \bibnamefont{and}
  \bibinfo{author}{\bibfnamefont{Y.}~\bibnamefont{Berthaud}},
  \bibinfo{journal}{Europhys. Lett.} \textbf{\bibinfo{volume}{28}},
  \bibinfo{pages}{585} (\bibinfo{year}{1994}).

\bibitem[{\citenamefont{Milman et~al.}(1993)\citenamefont{Milman, Blumenfeld,
  Stelmashenko, and Ball}}]{Milman93}
\bibinfo{author}{\bibfnamefont{V.~Y.} \bibnamefont{Milman}},
  \bibinfo{author}{\bibfnamefont{R.}~\bibnamefont{Blumenfeld}},
  \bibinfo{author}{\bibfnamefont{N.~A.} \bibnamefont{Stelmashenko}},
  \bibnamefont{and} \bibinfo{author}{\bibfnamefont{R.~C.} \bibnamefont{Ball}},
  \bibinfo{journal}{Phys. Rev. Lett.} \textbf{\bibinfo{volume}{71}},
  \bibinfo{pages}{204} (\bibinfo{year}{1993}).

\bibitem[{\citenamefont{Milman et~al.}(1994)\citenamefont{Milman, Stelmashenko,
  and Blumenfeld}}]{Milman94}
\bibinfo{author}{\bibfnamefont{V.~Y.} \bibnamefont{Milman}},
  \bibinfo{author}{\bibfnamefont{N.~A.} \bibnamefont{Stelmashenko}},
  \bibnamefont{and}
  \bibinfo{author}{\bibfnamefont{R.}~\bibnamefont{Blumenfeld}},
  \bibinfo{journal}{Prog. Mater. Sci.} \textbf{\bibinfo{volume}{38}},
  \bibinfo{pages}{425} (\bibinfo{year}{1994}).

\bibitem[{\citenamefont{McAnulty et~al.}(1992)\citenamefont{McAnulty, Meisel,
  and Cote}}]{McAnulty92}
\bibinfo{author}{\bibfnamefont{P.}~\bibnamefont{McAnulty}},
  \bibinfo{author}{\bibfnamefont{L.~V.} \bibnamefont{Meisel}},
  \bibnamefont{and} \bibinfo{author}{\bibfnamefont{P.~J.} \bibnamefont{Cote}},
  \bibinfo{journal}{Phys. Rev. A} \textbf{\bibinfo{volume}{46}},
  \bibinfo{pages}{3523} (\bibinfo{year}{1992}).

\bibitem[{\citenamefont{Bouchaud et~al.}(1993)\citenamefont{Bouchaud, Bouchaud,
  Lapasset, and Planès}}]{Bouchaud93}
\bibinfo{author}{\bibfnamefont{J.~P.} \bibnamefont{Bouchaud}},
  \bibinfo{author}{\bibfnamefont{E.}~\bibnamefont{Bouchaud}},
  \bibinfo{author}{\bibfnamefont{G.}~\bibnamefont{Lapasset}}, \bibnamefont{and}
  \bibinfo{author}{\bibfnamefont{J.}~\bibnamefont{Planès}},
  \bibinfo{journal}{Phys. Rev. Lett.} \textbf{\bibinfo{volume}{71}},
  \bibinfo{pages}{2240} (\bibinfo{year}{1993}).

\bibitem[{\citenamefont{Ertas and Kardar}(1992)}]{Ertas92}
\bibinfo{author}{\bibfnamefont{D.}~\bibnamefont{Ertas}} \bibnamefont{and}
  \bibinfo{author}{\bibfnamefont{M.}~\bibnamefont{Kardar}},
  \bibinfo{journal}{Phys. Rev. Lett.} \textbf{\bibinfo{volume}{66}},
  \bibinfo{pages}{929} (\bibinfo{year}{1992}).

\bibitem[{\citenamefont{Ertas and Kardar}(1993)}]{Ertas93}
\bibinfo{author}{\bibfnamefont{D.}~\bibnamefont{Ertas}} \bibnamefont{and}
  \bibinfo{author}{\bibfnamefont{M.}~\bibnamefont{Kardar}},
  \bibinfo{journal}{Phys. Rev. E} \textbf{\bibinfo{volume}{48}},
  \bibinfo{pages}{1703} (\bibinfo{year}{1993}).

\bibitem[{\citenamefont{Nakano et~al.}(1995)\citenamefont{Nakano, k.~Kalia, and
  Vashishta}}]{Nakano95}
\bibinfo{author}{\bibfnamefont{A.}~\bibnamefont{Nakano}},
  \bibinfo{author}{\bibfnamefont{R.}~\bibnamefont{k.~Kalia}}, \bibnamefont{and}
  \bibinfo{author}{\bibfnamefont{P.}~\bibnamefont{Vashishta}},
  \bibinfo{journal}{Phys. Rev. Lett.} \textbf{\bibinfo{volume}{75}},
  \bibinfo{pages}{3138} (\bibinfo{year}{1995}).

\bibitem[{\citenamefont{Calvet et~al.}(unpublished)\citenamefont{Calvet,
  Fisher, and Mandelbrot}}]{Mandelbrot2}
\bibinfo{author}{\bibfnamefont{L.}~\bibnamefont{Calvet}},
  \bibinfo{author}{\bibfnamefont{A.}~\bibnamefont{Fisher}}, \bibnamefont{and}
  \bibinfo{author}{\bibfnamefont{B.~B.} \bibnamefont{Mandelbrot}},
  \bibinfo{journal}{Cowles Foundation paper 1164}
  (\bibinfo{year}{unpublished}).

\bibitem[{\citenamefont{Kantelhardt et~al.}(2002)\citenamefont{Kantelhardt,
  Zschiegner, Koscielny-Bunde, Havlin, Bunde, and Stanley}}]{Kantelhardt02}
\bibinfo{author}{\bibfnamefont{J.~W.} \bibnamefont{Kantelhardt}},
  \bibinfo{author}{\bibfnamefont{S.~A.} \bibnamefont{Zschiegner}},
  \bibinfo{author}{\bibfnamefont{E.}~\bibnamefont{Koscielny-Bunde}},
  \bibinfo{author}{\bibfnamefont{S.}~\bibnamefont{Havlin}},
  \bibinfo{author}{\bibfnamefont{A.}~\bibnamefont{Bunde}}, \bibnamefont{and}
  \bibinfo{author}{\bibfnamefont{H.~E.} \bibnamefont{Stanley}},
  \bibinfo{journal}{Physica A} \textbf{\bibinfo{volume}{316}},
  \bibinfo{pages}{87} (\bibinfo{year}{2002}).

\bibitem[{\citenamefont{Muzy et~al.}(1991)\citenamefont{Muzy, Bacry, and
  Arneodo}}]{Muzy91}
\bibinfo{author}{\bibfnamefont{J.~F.} \bibnamefont{Muzy}},
  \bibinfo{author}{\bibfnamefont{E.}~\bibnamefont{Bacry}}, \bibnamefont{and}
  \bibinfo{author}{\bibfnamefont{A.}~\bibnamefont{Arneodo}},
  \bibinfo{journal}{Phys. Rev. Lett.} \textbf{\bibinfo{volume}{67}},
  \bibinfo{pages}{3515} (\bibinfo{year}{1991}).

\bibitem[{\citenamefont{O\'{s}wi\c{e}cimka
  et~al.}(1992)\citenamefont{O\'{s}wi\c{e}cimka, Kwapie\'{n}, and
  Dro\.{z}d\.{z}}}]{Oswiecimka06}
\bibinfo{author}{\bibfnamefont{P.}~\bibnamefont{O\'{s}wi\c{e}cimka}},
  \bibinfo{author}{\bibfnamefont{J.}~\bibnamefont{Kwapie\'{n}}},
  \bibnamefont{and}
  \bibinfo{author}{\bibfnamefont{S.}~\bibnamefont{Dro\.{z}d\.{z}}},
  \bibinfo{journal}{Phys. Rev. A} \textbf{\bibinfo{volume}{46}},
  \bibinfo{pages}{3523} (\bibinfo{year}{1992}).

\bibitem[{\citenamefont{Peng et~al.}(1994)\citenamefont{Peng, Buldyrev, Havlin,
  Simons, Stanley, and Goldberger}}]{Peng}
\bibinfo{author}{\bibfnamefont{C.~K.} \bibnamefont{Peng}},
  \bibinfo{author}{\bibfnamefont{S.~V.} \bibnamefont{Buldyrev}},
  \bibinfo{author}{\bibfnamefont{S.}~\bibnamefont{Havlin}},
  \bibinfo{author}{\bibfnamefont{M.}~\bibnamefont{Simons}},
  \bibinfo{author}{\bibfnamefont{H.~E.} \bibnamefont{Stanley}},
  \bibnamefont{and} \bibinfo{author}{\bibfnamefont{A.~L.}
  \bibnamefont{Goldberger}}, \bibinfo{journal}{Phys. Rev. E}
  \textbf{\bibinfo{volume}{49}}, \bibinfo{pages}{1685} (\bibinfo{year}{1994}).

\bibitem[{\citenamefont{Feder}(1988)}]{Feder88}
\bibinfo{author}{\bibfnamefont{J.}~\bibnamefont{Feder}},
  \emph{\bibinfo{title}{Fractals}} (\bibinfo{publisher}{Plenum Press, New
  York.}, \bibinfo{year}{1988}).

\bibitem[{\citenamefont{Bouchaud et~al.}(1990)\citenamefont{Bouchaud, Lapasset,
  and Plan\'{e}s}}]{Bouchaud90}
\bibinfo{author}{\bibfnamefont{E.}~\bibnamefont{Bouchaud}},
  \bibinfo{author}{\bibfnamefont{G.}~\bibnamefont{Lapasset}}, \bibnamefont{and}
  \bibinfo{author}{\bibfnamefont{J.}~\bibnamefont{Plan\'{e}s}},
  \bibinfo{journal}{Europhys. Lett.} \textbf{\bibinfo{volume}{13}},
  \bibinfo{pages}{73} (\bibinfo{year}{1990}).

\bibitem[{\citenamefont{Schmittbuhl et~al.}(1995)\citenamefont{Schmittbuhl,
  Schmitt, and Scholz}}]{Schmittbuhl95}
\bibinfo{author}{\bibfnamefont{J.}~\bibnamefont{Schmittbuhl}},
  \bibinfo{author}{\bibfnamefont{F.}~\bibnamefont{Schmitt}}, \bibnamefont{and}
  \bibinfo{author}{\bibfnamefont{C.}~\bibnamefont{Scholz}},
  \bibinfo{journal}{J. Geophys. Res.} \textbf{\bibinfo{volume}{100}},
  \bibinfo{pages}{5953} (\bibinfo{year}{1995}).

\bibitem[{\citenamefont{Krug}(1994)}]{Krug94}
\bibinfo{author}{\bibfnamefont{J.}~\bibnamefont{Krug}}, \bibinfo{journal}{Phys.
  Rev. Lett.} \textbf{\bibinfo{volume}{72}}, \bibinfo{pages}{2907}
  (\bibinfo{year}{1994}).

\bibitem[{\citenamefont{Bohr et~al.}(1992)\citenamefont{Bohr, Grinstein,
  Jayaprakash, Jensen, Krug, and Mukamel}}]{Bohr92}
\bibinfo{author}{\bibfnamefont{T.}~\bibnamefont{Bohr}},
  \bibinfo{author}{\bibfnamefont{G.}~\bibnamefont{Grinstein}},
  \bibinfo{author}{\bibfnamefont{C.}~\bibnamefont{Jayaprakash}},
  \bibinfo{author}{\bibfnamefont{M.~H.} \bibnamefont{Jensen}},
  \bibinfo{author}{\bibfnamefont{J.}~\bibnamefont{Krug}}, \bibnamefont{and}
  \bibinfo{author}{\bibfnamefont{D.}~\bibnamefont{Mukamel}},
  \bibinfo{journal}{Physica A} \textbf{\bibinfo{volume}{59D}},
  \bibinfo{pages}{177} (\bibinfo{year}{1992}).

\bibitem[{\citenamefont{Frisch}(1995)}]{Frisch}
\bibinfo{author}{\bibfnamefont{U.}~\bibnamefont{Frisch}},
  \emph{\bibinfo{title}{Turbulence : the legacy of A. N. Kolmogorov}}
  (\bibinfo{publisher}{Cambridge University Press}, \bibinfo{year}{1995}).

\bibitem[{\citenamefont{Bershadskii}(1997)}]{Bershadskii}
\bibinfo{author}{\bibfnamefont{A.}~\bibnamefont{Bershadskii}},
  \bibinfo{journal}{Europhys. Lett.} \textbf{\bibinfo{volume}{39}},
  \bibinfo{pages}{587} (\bibinfo{year}{1997}).

\bibitem[{\citenamefont{Kailasnath et~al.}(1992)\citenamefont{Kailasnath,
  Sreenivasan, and Stolovitzky}}]{Kailasnath92}
\bibinfo{author}{\bibfnamefont{P.}~\bibnamefont{Kailasnath}},
  \bibinfo{author}{\bibfnamefont{K.~R.} \bibnamefont{Sreenivasan}},
  \bibnamefont{and}
  \bibinfo{author}{\bibfnamefont{G.}~\bibnamefont{Stolovitzky}},
  \bibinfo{journal}{Phys. Rev. Lett.} \textbf{\bibinfo{volume}{68}},
  \bibinfo{pages}{2766} (\bibinfo{year}{1992}).

\bibitem[{\citenamefont{Mandelbrot}(1969)}]{Mandelbrot}
\bibinfo{author}{\bibfnamefont{B.}~\bibnamefont{Mandelbrot}},
  \emph{\bibinfo{title}{On Intermittent Free Turbulence, Turbulence of Fluids
  and Plasmas}} (\bibinfo{publisher}{Interscience, New York},
  \bibinfo{year}{1969}).

\bibitem[{\citenamefont{Ivanov et~al.}(1999)\citenamefont{Ivanov, Amaral,
  Goldberger, Havlin, Michael, Rosenblum, Struzikk, and Stanley}}]{Ivanov99}
\bibinfo{author}{\bibfnamefont{P.~C.} \bibnamefont{Ivanov}},
  \bibinfo{author}{\bibfnamefont{L.~A.~N.} \bibnamefont{Amaral}},
  \bibinfo{author}{\bibfnamefont{A.~L.} \bibnamefont{Goldberger}},
  \bibinfo{author}{\bibfnamefont{S.}~\bibnamefont{Havlin}},
  \bibinfo{author}{\bibnamefont{Michael}},
  \bibinfo{author}{\bibfnamefont{G.}~\bibnamefont{Rosenblum}},
  \bibinfo{author}{\bibfnamefont{Z.~R.} \bibnamefont{Struzikk}},
  \bibnamefont{and} \bibinfo{author}{\bibfnamefont{H.~E.}
  \bibnamefont{Stanley}}, \bibinfo{journal}{Nature}
  \textbf{\bibinfo{volume}{399}}, \bibinfo{pages}{461} (\bibinfo{year}{1999}).

\bibitem[{\citenamefont{Tanguy et~al.}(1998)\citenamefont{Tanguy, Gounelle, and
  Roux}}]{Tanguy98}
\bibinfo{author}{\bibfnamefont{A.}~\bibnamefont{Tanguy}},
  \bibinfo{author}{\bibfnamefont{M.}~\bibnamefont{Gounelle}}, \bibnamefont{and}
  \bibinfo{author}{\bibfnamefont{S.}~\bibnamefont{Roux}},
  \bibinfo{journal}{Phys. Rev. E} \textbf{\bibinfo{volume}{58}},
  \bibinfo{pages}{1577} (\bibinfo{year}{1998}).

\bibitem[{\citenamefont{Gao and Rice}(1989)}]{Gao89}
\bibinfo{author}{\bibfnamefont{H.}~\bibnamefont{Gao}} \bibnamefont{and}
  \bibinfo{author}{\bibfnamefont{J.~R.} \bibnamefont{Rice}},
  \bibinfo{journal}{J. Appl. Mech.} \textbf{\bibinfo{volume}{56}},
  \bibinfo{pages}{828} (\bibinfo{year}{1989}).

\bibitem[{\citenamefont{Calvet et~al.}(1997)\citenamefont{Calvet, Fisher, and
  Mandelbrot}}]{cowles}
\bibinfo{author}{\bibfnamefont{L.}~\bibnamefont{Calvet}},
  \bibinfo{author}{\bibfnamefont{A.}~\bibnamefont{Fisher}}, \bibnamefont{and}
  \bibinfo{author}{\bibfnamefont{B.~B.} \bibnamefont{Mandelbrot}}
  (\bibinfo{year}{1997}), \bibinfo{note}{cowles Foundation Discussion Paper No.
  1165}.

\end{thebibliography}
\end{document}